\documentclass{vgtc}                          

\graphicspath{{figures/}{pictures/}{images/}{./}} 
\usepackage{wrapfig}
\usepackage{times}                     

\usepackage{tabu}                      
\usepackage{booktabs}                  
\usepackage{lipsum}                    
\usepackage{mwe}                       
\usepackage{balance}
\usepackage{mathptmx}                  
\usepackage{amsmath}
\usepackage{paralist,bbding,pifont}
\usepackage{authblk}

\newcommand{\ouyangre}{\textcolor{black}}

\vgtccategory{Research}

\vgtcinsertpkg
\def\degree{${}^{\circ}$}

\title{OceanVive: An Immersive Visualization System for Communicating Complex Oceanic Phenomena}

\makeatletter

\renewcommand\AB@affilsepx{, \protect\Affilfont}
\makeatother
\author[1,2]{Yang Ouyang, Yuchen Wu\thanks{e-mail: \{ouyy,wuych3\}@shanghaitech.edu.cn (Equal contribution)}}
\author[1,2]{Xiyuan Wang, Laixin Xie\thanks{e-mail: \{wangxy7,xielx\}@shanghaitech.edu.cn}}
\author[2,3]{Weicong Cheng, Jianping Gan \thanks{e-mail: \{chengwc,magan\}@ust.hk}}

\author[1]{\\\ Quan Li\thanks{e-mail: liquan@shanghaitech.edu.cn (Corresponding author)}}
\author[2,3]{Xiaojuan Ma\thanks{e-mail: mxj@cse.ust.hk (Corresponding author)}}

\affil[1]{ShanghaiTech University}
\affil[2]{Center for Ocean Research in Hong Kong}
\affil[3]{Hong Kong University of Science and Technology}

\teaser{
    \centering
    \vspace{-6mm}
      \includegraphics[width=\linewidth]{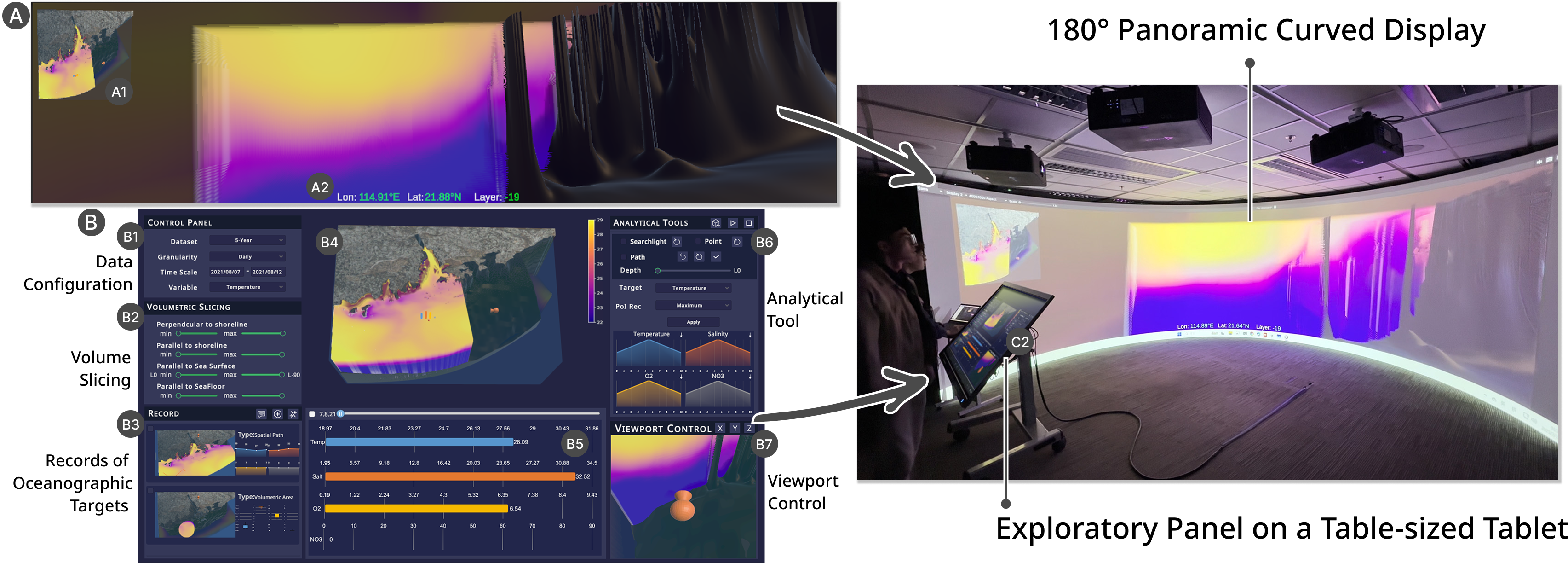}
      \vspace{-6mm}
      \caption{We demonstrate \textit{OceanVive} in an immersive environment: (A) 180\degree Panoramic Curved Display and (B) Exploratory Panel.}
      \vspace{-2mm}
      \label{fig:teaser}
}
\abstract{Communicating the complexity of oceanic phenomena—such as hypoxia and acidification—poses a persistent challenge for marine science. Despite advances in sensing technologies and computational models, conventional formats like static visualizations and text-based reports often fall short in conveying the dynamics of ocean changes. To address this gap, we present \textit{OceanVive}, an immersive and interactive visualization system that transforms complex ocean datasets into navigable spatial narratives. \textit{OceanVive} incorporates an exploratory panel on a table-sized tablet for managing immersive content on a large screen and integrates adaptive visual encodings, contextual storytelling, and intuitive navigation pathways to support effective communication. We validate the system through expert interviews, demonstrating its potential to enhance science communication and promote deeper public understanding.}    


\begin{document}
\firstsection{Introduction}

\maketitle

\par Human activities and climate change are exerting growing impacts on marine ecosystems, driving ecological disturbances such as hypoxia and ocean acidification~\cite{breitburg2018declining,gobler2016hypoxia}. Hypoxia results in low-oxygen ``dead zones''~\cite{diaz2008spreading} that severely compromise the survival of marine organisms, while the ocean's increased absorption of atmospheric $CO_2$ reduces seawater pH, threatening the health of coral reefs and shell-forming species~\cite{doney2009ocean}. Both phenomena are intensifying in frequency and severity, underscoring the urgent need for enhanced monitoring and a deeper understanding of their long-term ecological consequences~\cite{rabalais2010dynamics}.

\par Recent advances in ocean observation technologies~\cite{le2019observation,subramanian2019ocean} and computational modeling frameworks~\cite{zhao2024applications} have significantly enhanced our capacity to elucidate the mechanisms driving oceanic changes and to predict future trends. However, conventional communication formats such as scientific publications, static visualizations, and slide decks~\cite{franconeri2021science,tory2021finding} often inadequately convey the multidimensional nature of these changes. Text-centric reports and rigid visuals struggle to represent intricate spatial-temporal patterns, while existing dashboards and visualization systems prioritize data manipulation over intuitive storytelling~\cite{tory2021finding}. These shortcomings can impede stakeholders' ability to synthesize insights and implement timely, evidence-informed decisions critical for marine ecosystem protection~\cite{cvitanovic2015improving}.

\par To bridge this gap, immersive and interactive visualization tools are emerging as critical solutions~\cite{ren2024bridging,billen2008geoscience}. Mixed immersive approaches, which integrate dynamic spatial representations with user-driven interactions, offer two key advantages: 1) high interactivity and real-time responsiveness, which provide immediate feedback to user actions and environmental inputs, thereby enhancing engagement and cognitive processing; and 2) adaptive narrative structures with contextual continuity, allowing users to navigate content through personalized exploratory paths while preserving the coherence of the immersive experience.



\par This study presents a systematic approach to advancing oceanic science communication through immersive design. As a first step, we collaborated with five oceanographers to thoroughly explore current communication practices surrounding oceanic changes and to identify the design requirements for adopting advanced, immersive methods for more effective dissemination. \ouyangre{Our investigation revealed critical limitations in existing tools—ranging from static visualizations and non-interactive storytelling formats to analysis-centric platforms —that restrict accessibility and fail to support interactive, narrative-driven engagement for wider audiences.} Building on these insights, we developed \textit{OceanVive} — a visualization tool that transforms complex datasets into navigable spatial narratives. The system features an adaptive interface with dynamic visual encoding of key parameters, coupled with a streamlined exploratory panel for synchronized large-screen display. This adaptive framework seamlessly merges personalized data exploration with coherent storytelling, allowing users to navigate insights through intuitive pathways. 
\ouyangre{Unlike prior works, our system integrates story-driven visuals with real-time interactivity to make intricate oceanic narratives more accessible and engaging—empowering diverse audiences to understand complex ecological relationships while maintaining the rigor of scientific communication.}
Finally, we validated \textit{OceanVive}'s efficacy via semi-structured interviews with domain experts, demonstrating its potential to make marine science accessible to diverse audiences.

\section{Related Work}
\noindent\textbf{Interaction in Immersive Systems.} 
Large display systems are increasingly utilized for data visualization and collaborative analysis in immersive environments and science communication. However, enabling individualized understanding within these shared viewing contexts remains a key challenge. Some studies have addressed this by integrating auxiliary devices. Reipschlager et al.~\cite{reipschlager2020personal} used AR headsets to provide personalized overlays, while Letondal et al.~\cite{letondal2019exploring} explored touch-based flight control panels, proposing robust gestures that require minimal visual focus and leverage spatial and proprioceptive skills. Additionally, Chen et al.~\cite{chen2021photo4action} leveraged smartphones to facilitate independent exploration. \ouyangre{Flexible interaction methods are essential for engaging users in immersive environments. Researchers have explored alternatives to traditional gesture tracking and touch surfaces, such as using smartphones as pointers~\cite{siddhpuria2018pointing} or leveraging mobile cameras for multimodal tracking~\cite{babic2022understanding}. In tiled displays, spatially-aware devices~\cite{langner2018towards} and photo-based interaction~\cite{chen2021photo4action} enable content navigation through captured screen images. Submerse~\cite{boorboor2023submerse} further demonstrated the use of augmented reality (AR) for immersive storm surge simulations. Building on these approaches, our system employs a compact control panel—drawing inspiration from aircraft cockpit interfaces~\cite{ohlander2017user}—to manage immersive ocean content on a surrounding large display. Unlike prior AR/VR-dependent solutions, our room-scale setup delivers an accessible, shared immersive experience without requiring head-mounted devices.}

\noindent\textbf{View-Finding and Navigation in Immersive Environments.} Early work by Kamada and Kawai~\cite{kamada1988simple} focused on optimizing line projections to determine optimal views for basic shapes. Bonaventura et al.~\cite{bonaventura2018survey} provided an overview of viewpoint selection methods for polygonal data, while information entropy~\cite{feixas1999information} emerged as a metric for visible information. Arbel et al.~\cite{arbel1999viewpoint} applied Shannon entropy to monochrome object recognition. In the context of landscape scenes, Stoev et al.~\cite{stoev2002case} developed a method for automatic camera positioning based on maximizing scene depth and projected area for terrain datasets. Huang et al.~\cite{huang2016trip} introduced a camera motion design method for urban scenes, where a set of viewpoints is selected based on visible buildings' number and shapes. Lin et al.~\cite{lin2022capturing} proposed a comprehensive framework for large-scale scene perception and reconstruction. Submerse~\cite{boorboor2023submerse} introduced a dual-purpose view-finding algorithm that selects viewpoints maximizing information visibility while considering surrounding context and display configurations. Building upon these findings, our work enhances information visibility on the large screen for navigation path display while allowing free perspective switching on a tablet. \ouyangre{Navigation in immersive environments is essential for exploration, enabling users to traverse 3D spaces and engage with narrative content. While techniques such as gesture control~\cite{tursunov2024creating}, alternative inputs~\cite{swidrak2024beyond}, spatial manipulation~\cite{dong2021tailored}, and multi-scale systems~\cite{mirhosseini2019exploration} have been extensively studied—summarized by Luca et al.~\cite{di2021locomotion} in the Locomotion Vault—mainstream platforms still prioritize natural walking and teleportation for their simplicity and accessibility. In narrative contexts, navigation can be active or passive: active methods enhance interactivity but may cause disorientation, while passive approaches maintain narrative control at the cost of reduced user agency and potential motion sickness. This trade-off remains a central design challenge. Lages and Bowman~\cite{lages2018move} further showed that navigation performance depends on users’ spatial abilities. In our work, we explore a hybrid approach combining passive navigation with predefined targets, guided by a streamlined exploratory panel synchronized with a large-screen display.}

\noindent\textbf{Visualization for Ocean Data.} Researchers have developed diverse techniques to visualize complex, multidimensional datasets, including 2D and 3D methods that combine temperature, salinity, density surfaces, and bathymetry models~\cite{head1997applications}. These techniques are useful for quality control, feature detection, and analyzing both field-collected and simulated data, often leveraging animation and computer vision technologies~\cite{Rosenblum1989VisualizingOD}. More recently, InfoVis techniques, such as 2D/3D charts, surface maps, and choropleth maps, have been employed to analyze seasonal patterns in wind and wave data from oceanographic buoys~\cite{Monteiro2019VisualAS}. While previous efforts focused on data analysis, our work emphasizes effective communication within an immersive environment. We aim to employ visualizations that are not only validated by collaborating experts but also readily comprehensible to the audience.

\section{Observational Study}
\par We collaborated closely with five oceanographers (\textbf{E1–E5}) from a local oceanography institute, each with extensive experience in communicating ocean phenomena to peers, the general public, and stakeholders. This study aims to uncover current practices and identify bottlenecks in ocean communication. To achieve this, we conducted individual semi-structured interviews, each lasting approximately 30 minutes, focusing on their methods for grounding, curating, and delivering information. In addition, we held in-depth discussions to further explore challenges encountered across different stages of the communication process. Building upon the insights gained from interviews, we identified five design requirements reflecting the needs and challenges oceanographers faced:

\par \textbf{R1. Enable flexible and intuitive observation of ocean data.}
Oceanographers emphasized the need for more adaptable and intuitive methods to inspect ocean data. \textbf{E2} noted that raw datasets (e.g., satellite imagery, buoy measurements) often lack the contextual clarity needed by stakeholders, while \textbf{E3} pointed out that static sectional views fail to capture adjacent dynamics critical for understanding phenomena such as algal blooms. A system should thus support intuitive exploration and flexible slicing to better facilitate data observation.

\par \textbf{R2. Support concurrent inspection of diverse variables.}
Experts highlighted the interdependence of variables like sea surface temperature, salinity, and chlorophyll concentration. \textbf{E1} explained, ``\textit{Isolating a single variable risks oversimplification—for example, understanding hypoxia events requires correlating oxygen levels with nutrient runoff and stratification.}'' A system should enable synchronized overlays or side-by-side comparisons, complemented by automated statistical summaries (e.g., anomalies, trends) to reveal causal relationships. \textbf{E4} further recommended incorporating anomaly detection algorithms to flag critical outliers.

\par \textbf{R3. Accommodate heterogeneous characterization of oceanographic entities.} Ocean phenomena differ greatly in scale and structure, from discrete eddies to basin-wide currents. \textbf{E3} observed that current tools often impose uniform representations, masking distinct behaviors. A solution should support entity-specific visual encodings—such as points for drain outlets, paths for currents, and volumes for pollutant dispersion.

\par \textbf{R4. Ensure continuous and contextual delivery of ocean dynamics.} Stakeholders often need real-time updates alongside holistic contextualization. \textbf{E2} remarked, ``\textit{For hurricane events, we typically rely on predefined materials, but policymakers may shift focus or raise new questions.}'' A system should maintain continuous narrative coherence throughout the communication process to support situational understanding. \textbf{E1} also emphasized the value of immersive observations—such as viewing phenomena within the water body—to enhance credibility.

\par \textbf{R5. Balance semi-automation with human autonomy.}
While automation enhances efficiency, experts cautioned against over-reliance. \textbf{E4} warned, ``\textit{Algorithms might miss region-specific nuances known only to local experts.}'' Systems should offer template-based workflows (e.g., point-path comparisons) for drafting observations, while allowing experts to adjust perspectives and narrative emphasis as needed.

\section{OceanVive}
\par Building on these requirements, we developed \textit{OceanVive}, a hybrid system designed to enhance the communication of oceanographic phenomena through a ``submarine'' metaphor. The system integrates an \textit{Exploratory Dashboard} for overview and data manipulation, alongside an \textit{Immersive Viewport} for in-depth data exploration, facilitating more intuitive and effective communication. Prior studies have shown that CAVEs and immersive displays improve data realism and support more efficient exploration of three- and higher-dimensional datasets by engaging peripheral vision~\cite{laha2012effects} (\textbf{R1}, \textbf{R4}). Moreover, Beaudouin-Lafon~\cite{beaudouin2011lessons} emphasized the advantages of large shared displays for collaborative data analysis, complemented by personal devices for focused, individual tasks. Inspired by these findings, \textit{OceanVive} deploys visual communication across a concave wall-sized display for shared context and a table-sized tablet for detailed, fine-grained interactions.

\subsection{Data Characteristics and Processing}
\par The ocean data were collected from the Great Bay Area in the China Seas through in situ measurements and provided by domain experts. The dataset covers a five-year period (2019–2024), with daily records of four oceanographic variables: temperature, salinity, dissolved oxygen, and nitrate concentration. These data are presented in a three-dimensional grid (lattice) format ($0 \sim -90$ layers from surface to seafloor; each layer is a $400 \times 441$ grid), incorporating a mask that represents the continental shelf topography.

\par Using this data mask as a reference, we first modeled the Bay Area's coastal terrain for Unity asset development. The scale ratio was calculated from the variable data, and, in line with expert guidelines, each 3D value was assigned its corresponding color. The finalized data were then packaged as Unity-compatible assets.

\subsection{System Walkthrough}

\ouyangre{We illustrate a typical system workflow with \textit{OceanVive}, demonstrating how users navigate, analyze, and present oceanographic phenomena across two coordinated surfaces: the \textit{Exploratory Panel} on a table-size tablet (\autoref{fig:teaser}-C2) and the \textit{Immersive Viewport} on a wall-size curved display (\autoref{fig:teaser}-C1).}


\textbf{Data Configuration and Initial Exploration.} The walkthrough begins with Data Configuration in the Control Panel (\autoref{fig:teaser}-B1), where users select the dataset, temporal granularity, time range, and up to four variables of interest. These settings can be dynamically adjusted during exploration through dropdown menus and input fields, establishing the foundational context for subsequent tasks. \ouyangre{The color scheme employed in the current visualizations adheres to conventions widely recognized by domain experts and has been refined through their direct feedback and guidance.} Following configuration, the system renders a 3D Ocean Miniature (\autoref{fig:teaser}-B4) that visualizes bathymetry and data volume using interpolated volume rendering. Users can interactively zoom, rotate, and pan the miniature, enabling exploration of spatial structures. This intuitive overview facilitates user orientation and the early formation of hypotheses  (\textbf{R1}, \textbf{R4}).

\begin{figure}[h]
    \centering
     \vspace{-2mm}
    \includegraphics[width=\linewidth]{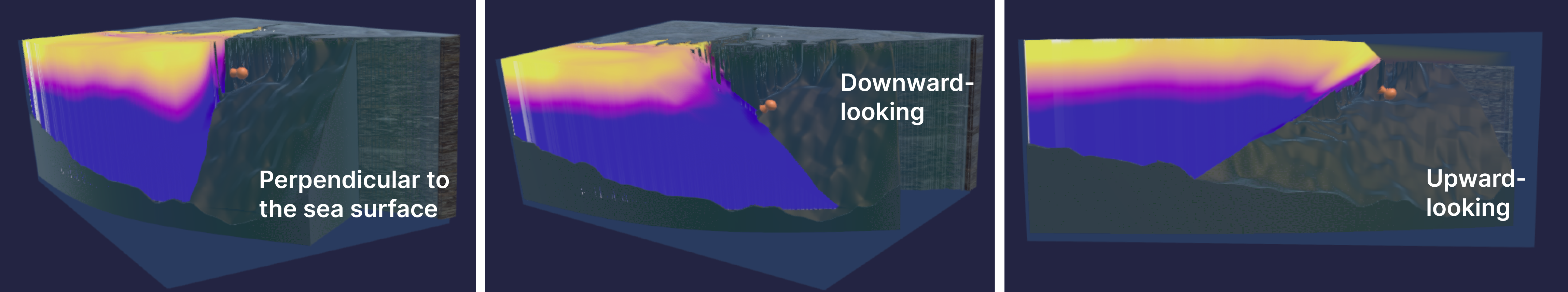}
    \vspace{-6mm}
    \caption{Real-time dynamic slicing method enables flexible, contextual, and immediate observation of data from any angle and position.}
    \vspace{-3mm}
    \label{fig:dynamic_slicing}
\end{figure}

\textbf{Slicing for In-Depth Inspection.} For fine-grained observation, users can activate \textit{Volumetric Slicing} (\autoref{fig:teaser}-B2), selecting from standardized slicing schemes (e.g., perpendicular to shoreline, parallel to surface). Adjustable sliders define slice bounds, instantly updating the cross-sections on the miniature (\autoref{fig:dynamic_slicing}). Simultaneously, the \textit{Viewport Control} (\autoref{fig:teaser}-B7) enables embodied navigation of the immersive viewport. Users swipe to adjust orientation, snap to XYZ axes via buttons, and observe a vertical dynamic slice aligned with the current viewing direction. The function supports contextual, continuous exploration of the data volume (\textbf{R1}, \textbf{R4}).

\begin{figure}[h]
    \centering
     \vspace{-2mm}
    \includegraphics[width=\linewidth]{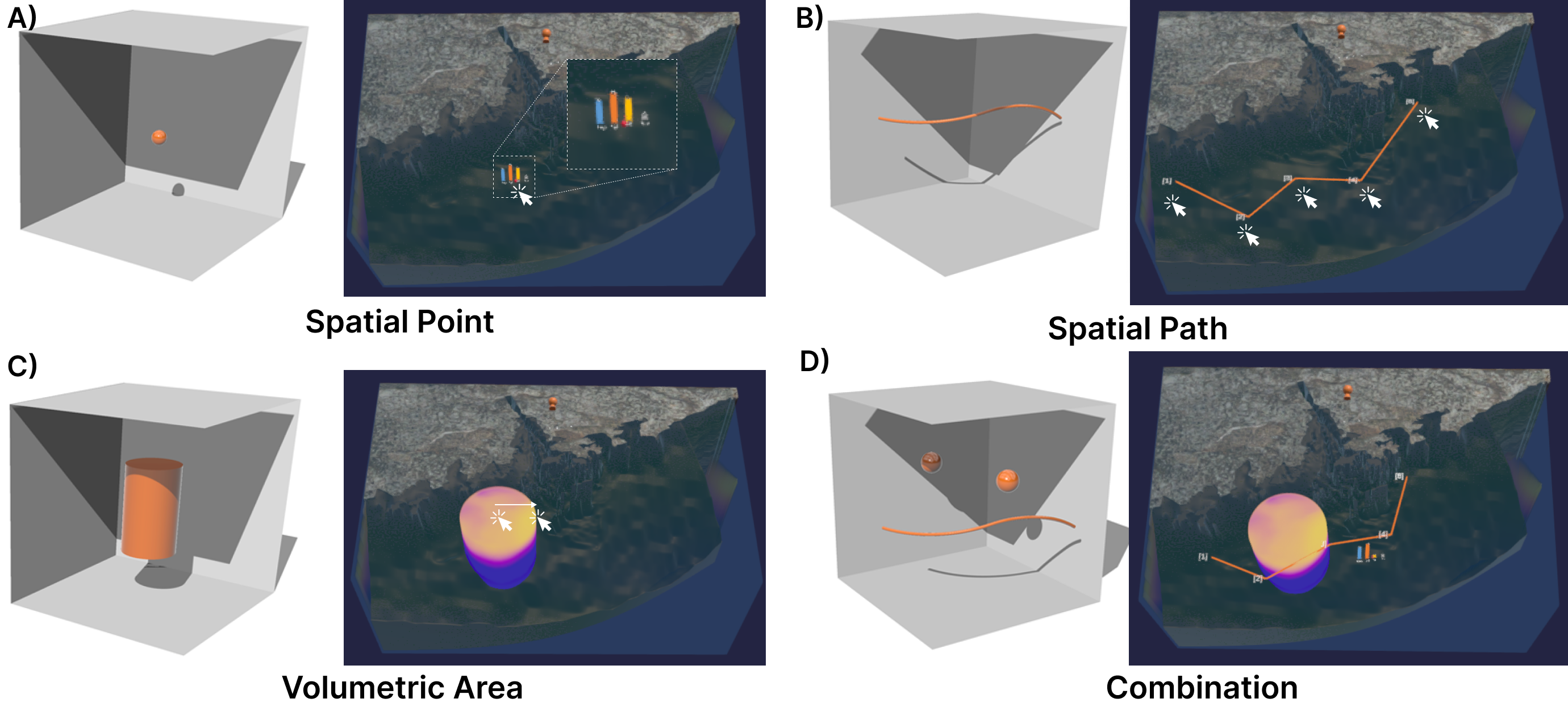}
    \vspace{-6mm}
    \caption{Four observational targets in OceanVive, including A) Spatial Point, designated by clicking on the miniature, B) Spatial Path, designated by connecting a series of spatial points, C) Volumetric Area, designated by circle selection on the miniature, and D) the combination of multiple targets.}
    \vspace{-3mm}
    \label{fig:targets}
\end{figure}

\textbf{From Exploration to Target Specification.} After navigating to a region of interest, users switch to \textit{Analytical Tools} to designate observational targets directly on the miniature (\autoref{fig:targets}). It provides a comprehensive set of core functionalities to enable the in situ designation of oceanographic entities directly on the ocean miniature. After consulting with oceanographers, we have identified three heterogeneous types of oceanographic entities as observational targets (\textbf{R3}), namely \textit{Spatial Point}, \textit{Spatial Path}, and \textit{Volumetric Area} (or \textit{Searchlight}). As depicted in \autoref{fig:targets}A, a \textit{Spatial Point} is a point located spatially within and on the miniature, represented as a red dot. The \textit{Spatial Path} (\autoref{fig:targets}B) is defined by a series of spatial points within the miniature and is visualized as an orange polyline. The \textit{Volumetric Area} (\autoref{fig:targets}C) represents a 3D area with specified height and position, rendered as a cylindrical volume. Additionally, \textit{OceanVive} enables users to combine multiple targets for observation (\autoref{fig:targets}D).



\par \textbf{From Spatial Points to Navigation Paths.} As shown in the upper part of \autoref{fig:teaser}-B6, through checking the checkboxes, users can designate \textit{Spatial Point} and \textit{Spatial Path} by clicking on the top surface of the miniature and adjust point depth by adjusting the depth slider below. For \textit{Volumetric Area}, users can perform the first click on the surface to designate the center and the second click to designate the radius. Users can adjust the height and vertical position of the area by adjusting the \textit{Parallel to sea surface} min-max slider. Through pressing the \raisebox{-0.5ex}{\includegraphics[height=2.2ex]{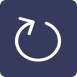}} button beside, users can reset the corresponding targets in the scene. The \raisebox{-0.5ex}{\includegraphics[height=2.2ex]{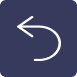}} and \raisebox{-0.5ex}{\includegraphics[height=2.2ex]{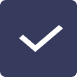}} buttons enable the retraction of the last path point and the completion of the entire path. \ouyangre{In the Analytical Tools section, \textit{OceanVive} adopts a bottom-up, data-driven recommendation approach to support data grounding. Users select target variables and a Position-of-Interest (POI) model—such as \textit{Maximum}, \textit{Minimum}, \textit{Temporal Gradient}, or \textit{Spatial Gradient}—via dropdown menus. Upon clicking ``Apply'', the system highlights recommended POIs as green points within the Miniature view.}

\par
\begin{wrapfigure}{r}{0.5\columnwidth}
\vspace{-3mm}
    \centering
    \hspace*{-20pt} 
    \begin{minipage}{0.57\columnwidth}
    \includegraphics[width=\columnwidth]{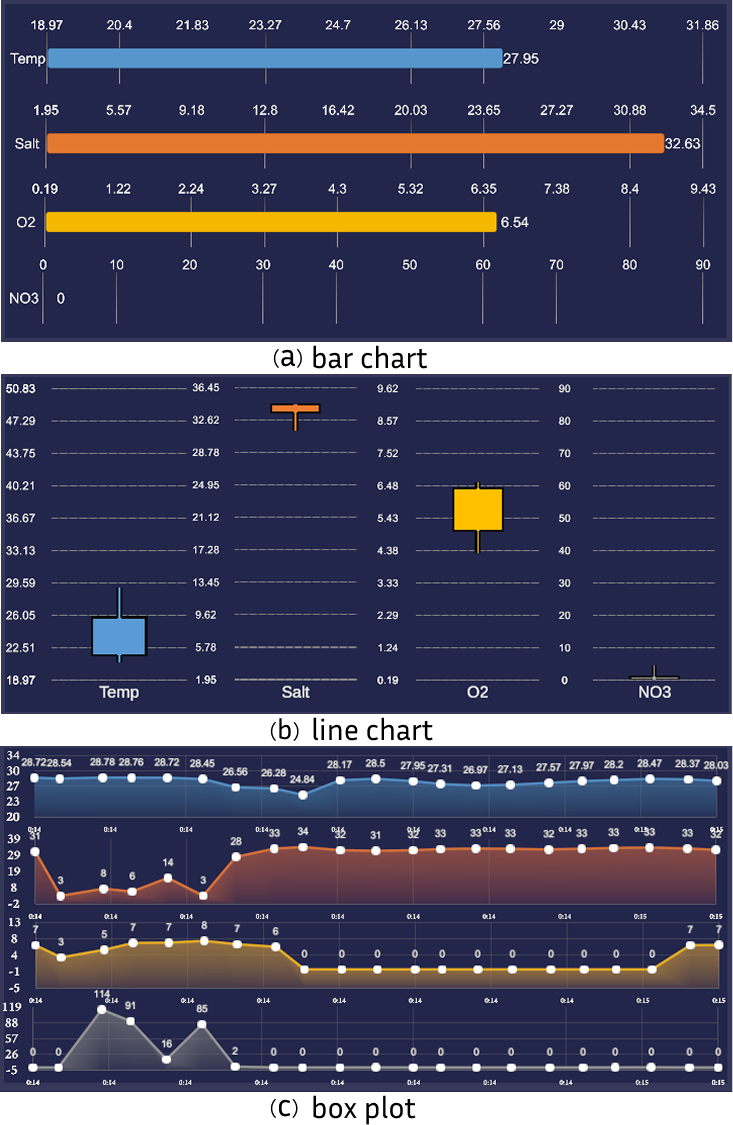}
    \vspace{-6mm}
    \caption{Three switchable charts tailored to current observational target: bar chart, line chart, and box plot.}
    \end{minipage}
    \vspace{-7mm}
    \label{fig:switchable_charts}
\end{wrapfigure}

 Each designated target triggers corresponding visualizations in the \textit{Switchable Charts} (\autoref{fig:switchable_charts}). Depending on the target type, the system presents: Bar charts for \textit{Spatial Point}, Line charts for \textit{Spatial Path}, Box plots for \textit{Volumetric Area}. These visualizations support concurrent inspection of four variables (\textbf{R2}). A \textit{Timeline} above the charts enables playback across time, with slider interaction to control progression. Temporal interpolation ensures continuous rendering during animated exploration (\textbf{R4}).

\begin{figure}[h]
\vspace{-3mm}
    \centering
    \includegraphics[width=\linewidth]{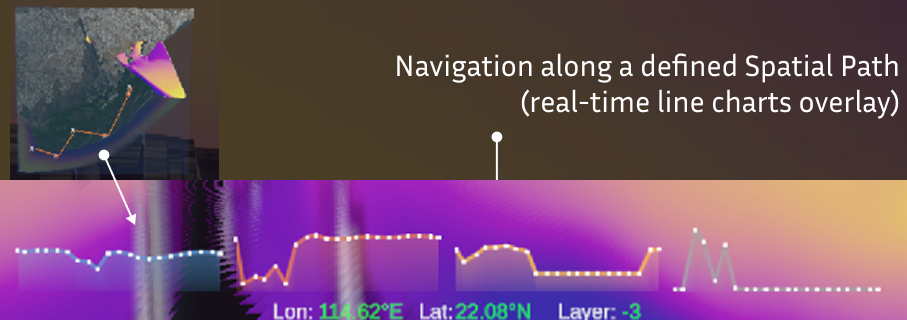}
    \vspace{-6mm}
    \caption{Four overlying line charts below indicate the sampled values along the route.}
    \label{fig:immerisve_display}
\end{figure}

\textbf{Target Recording and Contextual Presentation with Immersive Viewport.} \ouyangre{Users curate findings through the Record module, where designated targets can be added to a record list by clicking the \raisebox{-0.5ex}{\includegraphics[height=2.2ex]{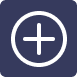}} button. Each entry includes a thumbnail, target type, and variable summaries, supporting structured documentation of exploration. For retrospective analysis and comparative storytelling, users can reconstruct and juxtapose recorded targets by selecting them and pressing the \raisebox{-0.5ex}{\includegraphics[height=2.2ex]{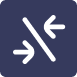}} button. Subsequently, users transition to the Immersive Viewport (\autoref{fig:immerisve_display}) for contextualized, shared presentation. This first-person visualization displays the viewport’s spatial location, accompanied by a mini-map and coordinate indicators for orientation. As users navigate along a defined Spatial Path, real-time line charts overlay beneath the main display to depict variable changes, streamlining interpretation without the need to alternate between devices.}



\vspace{-6mm}
\section{Expert Evaluation}
\ouyangre{We conducted an evaluation with five oceanographers (including three new participants), each with over five years of experience in ocean data analysis and communication. Participants completed a 15-minute tutorial using example datasets, followed by 15–20 minutes of hands-on exploration and a 15–20 minute semi-structured interview.} Feedback was categorized into three areas: \textit{tool perception}, \textit{workflow adaptation}, and \textit{concerns about the immersive experience}.  

\par \textbf{\textit{Perception of OceanVive}}. Experts generally viewed our approach favorably, highlighting its ability to foster intuitive understanding of complex ocean dynamics. Compared to tools they previously used, our system's intuitive, user-friendly design and innovative interactions enhance data communication, surpassing conventional tools by broadening perspectives and unlocking new analytical possibilities. 
Additionally, they appreciated the system's capacity to bridge scientific knowledge with public engagement, especially through its spatially anchored, immersive visualizations.

\par \textbf{\textit{Adaptation of workflows.}} Experts found the system's step-by-step workflow effective for exploring ocean data, offering an accessible yet adaptable authoring experience. This design lowers the barrier to engaging with complex datasets while maintaining flexibility for expert-level customization. They viewed it as a meaningful step toward enhancing the communication of ocean science, especially in interdisciplinary and educational contexts. \ouyangre{Experts also valued the streamlined, semi-automated recommendation algorithm, which supports data-driven analysis by enabling parameter adjustments to focus exploration and efficiently surface relevant patterns. However, some noted that the current approach differs from their hands-on analytical workflows. They suggested integrating domain-specific knowledge into the recommendation process to better align with expert reasoning and improve usability.} Additionally, several participants highlighted the system's potential to support collaboration, noting that its structured workflow could benefit other scientific fields that require transparent reasoning and reproducible processes.

\par \textbf{\textit{Concerns regarding the immersive experience.}} Several concerns were raised regarding the potential for misinterpretation, which could stem from the abstract nature of the visual elements. The complexity and ambiguity of these elements may challenge users' ability to fully grasp the intended message. In addition, there are worries about cognitive overload, as the immersive environment, combined with multi-sensory stimuli, could overwhelm users and hinder their ability to process information effectively. Some experts argue that immersive visualizations are effective, though there are times when switching back to static visualizations allows for a lighter and more straightforward way to observe data.

\section{Discussion, Limitation and Conclusion}
\par We propose \textit{OceanVive}, a tool designed to support the immersive dissemination and storytelling of complex oceanic changes. \textit{OceanVive} is highly customizable and modular, allowing it to be extended to a wide range of scientific tasks beyond oceanographic data. Researchers can tailor visualization strategies and communicative structures to meet domain-specific needs, enabling effective communication of insights derived from high-dimensional data. This flexibility makes it applicable not only to oceanography but also to fields such as genomics and materials science. In addition, beyond traditional data analysis, \textit{OceanVive} prioritizes the communication of findings, which experts recognize as crucial for externalizing, validating, and refining discoveries in collaborative scientific contexts. \ouyangre{While its semi-automated features streamline pattern identification, experts retain control to incorporate domain-specific knowledge, ensuring exploratory paths align with their reasoning.} Future work could enhance the system by integrating pre-set templates or event-driven functions, allowing users to annotate anticipated patterns in advance while seamlessly supporting both analysis and communication.

\par Our work has several limitations. First, while effective in our evaluations, the system's complexity creates a steep learning curve for novice users. Future iterations could streamline the user interface to improve usability. Additionally, its current reliance on traditional digital interfaces limits its adaptability and integrating VR/AR headset support in future versions could significantly broaden its scope. Finally, the absence of a user study with a diverse participant pool means we currently lack data on the system's performance across different user groups. Future research should address this gap to better understand its utility in varied contexts.

\section{Acknowledgments}
\par This work is supported by CORE, the joint research center for ocean research established between Laoshan Laboratory and the Hong Kong University of Science and Technology (HKUST); the National Natural Science Foundation of China (Grant No. 62372298); and the MoE Key Laboratory of Intelligent Perception and Human-Machine Collaboration (KLIP-HuMaCo).

\bibliographystyle{abbrv-doi}
\balance
\bibliography{template}
\end{document}